\documentclass[journal,twoside,final]{IEEEtran}
\usepackage[utf8]{inputenc}
\usepackage[T1]{fontenc}
\usepackage{graphicx}
\usepackage{amssymb}
\usepackage[boxed,noline]{algorithm2e}
\usepackage{multirow}
\usepackage{setspace}
\IEEEoverridecommandlockouts
\graphicspath{{./}}	%% specify a directory for figures

\title{On the Excess Bandwidth Allocation in ISP Traffic Control for Shared Access Networks}

\author{Kyeong Soo Kim, \IEEEmembership{Member, IEEE}\\
  \thanks{K. S. Kim is with the College of Engineering, Swansea University,
    Swansea, SA2 8PP, Wales United Kingdom (e-mail: k.s.kim@swansea.ac.uk).}%%
}

\begin{document}

\maketitle

\begin{abstract}
  Current practice of shaping subscriber traffic based on token bucket by
  Internet service provider (ISP) allows short-term fluctuations in its shaped
  rate and thereby enables a subscriber to transmit traffic at a higher rate
  than a negotiated long-term average. The traffic shaping, however, results in
  significant waste of network resources, especially when there are only a few
  active subscribers, because it cannot allocate excess bandwidth to active
  subscribers in the long term. In this letter we investigate the long-term
  aspect of resource sharing in ISP traffic control for shared access
  networks. We discuss major requirements for the excess bandwidth allocation in
  shared access networks and propose ISP traffic control schemes based on
  core-stateless fair queueing (CSFQ) and token bucket meters. Simulation
  results demonstrate that the proposed schemes allocate excess bandwidth among
  active subscribers in a fair and efficient way, while not compromising the
  service contracts specified by token bucket for conformant subscribers.
\end{abstract}


\section{Introduction}
\label{sec-1}

\IEEEPARstart{T}{he} practice of shaping subscriber traffic by Internet service
provider (ISP) has been under intensive study; for example, the effect of ISP
traffic shaping on various packet-level
\cite{bauer11:_power,sundaresan11:_broad_inter_perfor} and user-perceived
\cite{Kim:12-2} performances has been investigated, which provides a new insight
into the actual performance of broadband access networks.

One critical issue is that traffic shaping cannot allocate excess bandwidth to
active subscribers in the long term. This is because the traffic shaper based on
token bucket cannot take into account the status of other subscribers. As
extensively studied in \cite{bauer11:_power,Kim:12-2}, a large-size token bucket
enables sharing of excess bandwidth among active subscribers, but only in the
short period of time corresponding to the token bucket size.

The modification of token bucket algorithm to allocate excess bandwidth has been
studied in the context of fair queueing/scheduling
\cite{abendroth06:_solvin,kidambi99:_dynam_dtb} and differentiated services
(DiffServ) networks \cite{park03:_adapt_diffs}. The results of these studies,
however, cannot be applicable to the current ISP traffic control which is not
based on DiffServ. Also, the modification of token bucket algorithm and/or the
change of its negotiated parameters during the operation may raise the issue of
traffic conformance --- which is currently based on the original token bucket
algorithm --- and compromise the quality of service (QoS) of conformant traffic
as a result.

A desirable alternative to the traffic shaping based on a modified or adaptive
token bucket would be the use of the original token bucket as a meter in order
to separate traffic from a subscriber into conformant and non-conformant one and
treat them differently in further processing based on the status of a network
and other subscribers (e.g.,
\cite{huang06:_suppor,patt-shamir08:_compet_sla}). The issue of per-subscriber
allocation of excess bandwidth proportional to its negotiated long-term average
rate, however, has not been studied in this context.

In this letter we discuss major requirements for the excess bandwidth allocation
in shared access networks and propose ISP traffic control schemes based on
core-stateless fair queueing (CSFQ) \cite{stoica03:_core} and token bucket
meters that can meet the requirements.
\section{Excess Bandwidth Allocation}
\label{sec-2}
\subsection{Requirements}
\label{sec-2-1}

We define the excess bandwidth in downstream at time $t$ for an access network
with $N$ subscribers as follows:
\begin{equation}
C_{ex}(t) \triangleq C - r_{c}(t) ,
\label{eq:ex_bw}
\end{equation}
where $C$ is the capacity of the access link and $r_{c}(t)$ is the arrival rate
of conformant packets for all the subscribers from the network.\footnote{The discussions in this letter are also applicable to upstream traffic
  with minor modifications because the upstream traffic control in shared access
  networks is also centralized and located in the access switch (e.g., using
  grants in cable Internet and Ethernet passive optical network (EPON)).
 }
Below we set two major requirements that any excess bandwidth allocation schemes
should meet:
\begin{itemize}
\item The allocation of excess bandwidth should not compromise the QoS of
  subscribers' traffic conformant to service contracts based on the original
  token bucket algorithm.
\item Excess bandwidth should be allocated among active subscribers proportional to
  their negotiated long-term average rates, i.e., token generation rates.
\end{itemize}

The first requirement is more fundamental than the second one because both
subscribers and ISPs consider the excess bandwidth allocation as an optional
feature and therefore its benefit should not come at the expense of other
subscribers; note that the traffic conformance is solely based on the ISP
traffic control at the edge of the network and covers access links only.
The second requirement, on the other hand, enables ISPs to provide new service
and pricing schemes with more incentives to subscribers willing to pay more for
higher long-term average rates.
\subsection{ISP Traffic Control Schemes based on WFQ and CSFQ}
\label{sec-2-2}

\begin{figure}[!t]
\centering
\includegraphics[angle=-90,width=.85\linewidth]{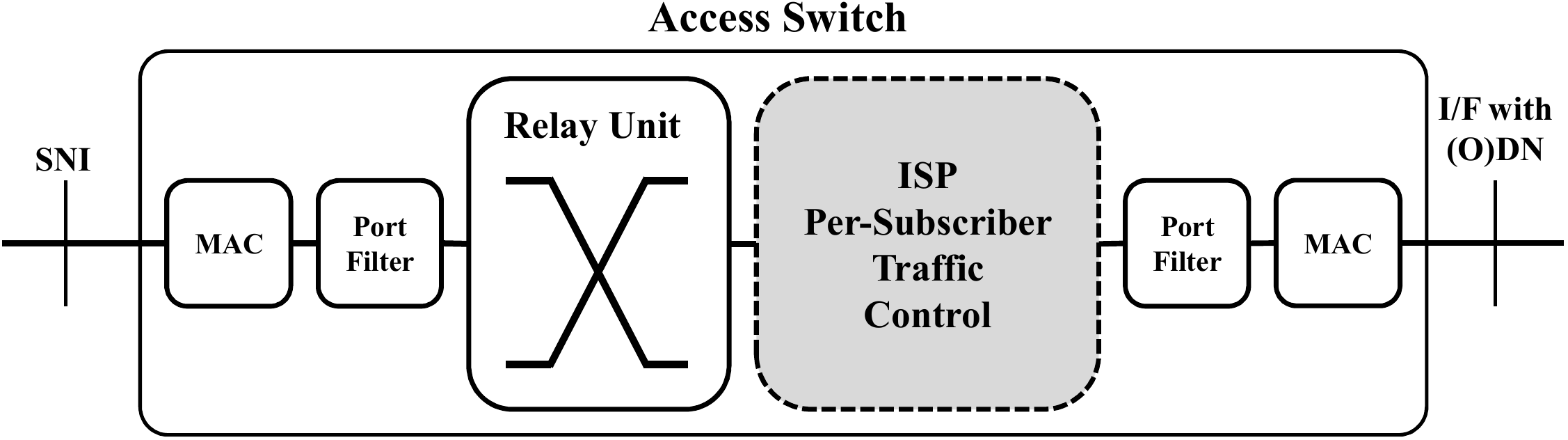}
\caption{Block diagram of an access switch for a shared access network.}
\label{fg:access_switch}
\end{figure}

\begin{figure}[!t]
\begin{center}
\includegraphics[angle=-90,width=0.7\linewidth]{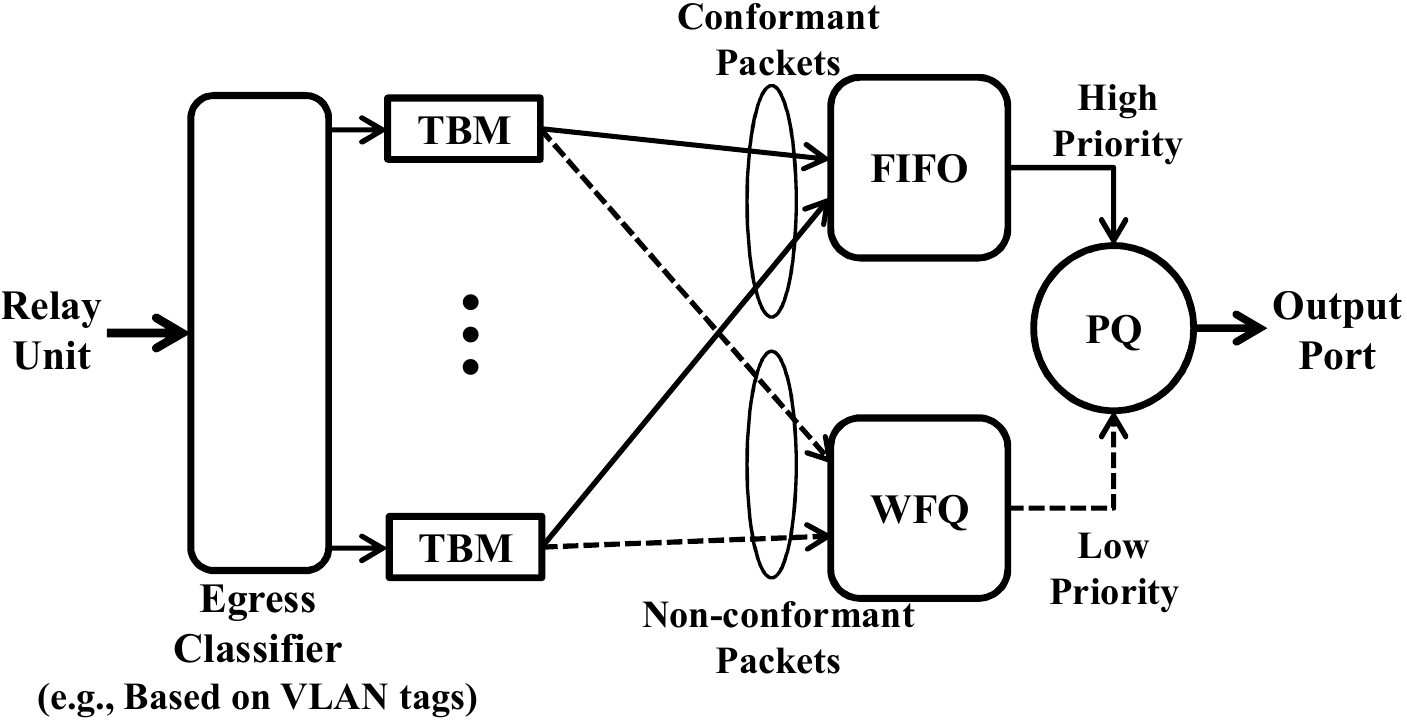} \\
{\scriptsize (a)} \\
\includegraphics[angle=-90,width=.85\linewidth]{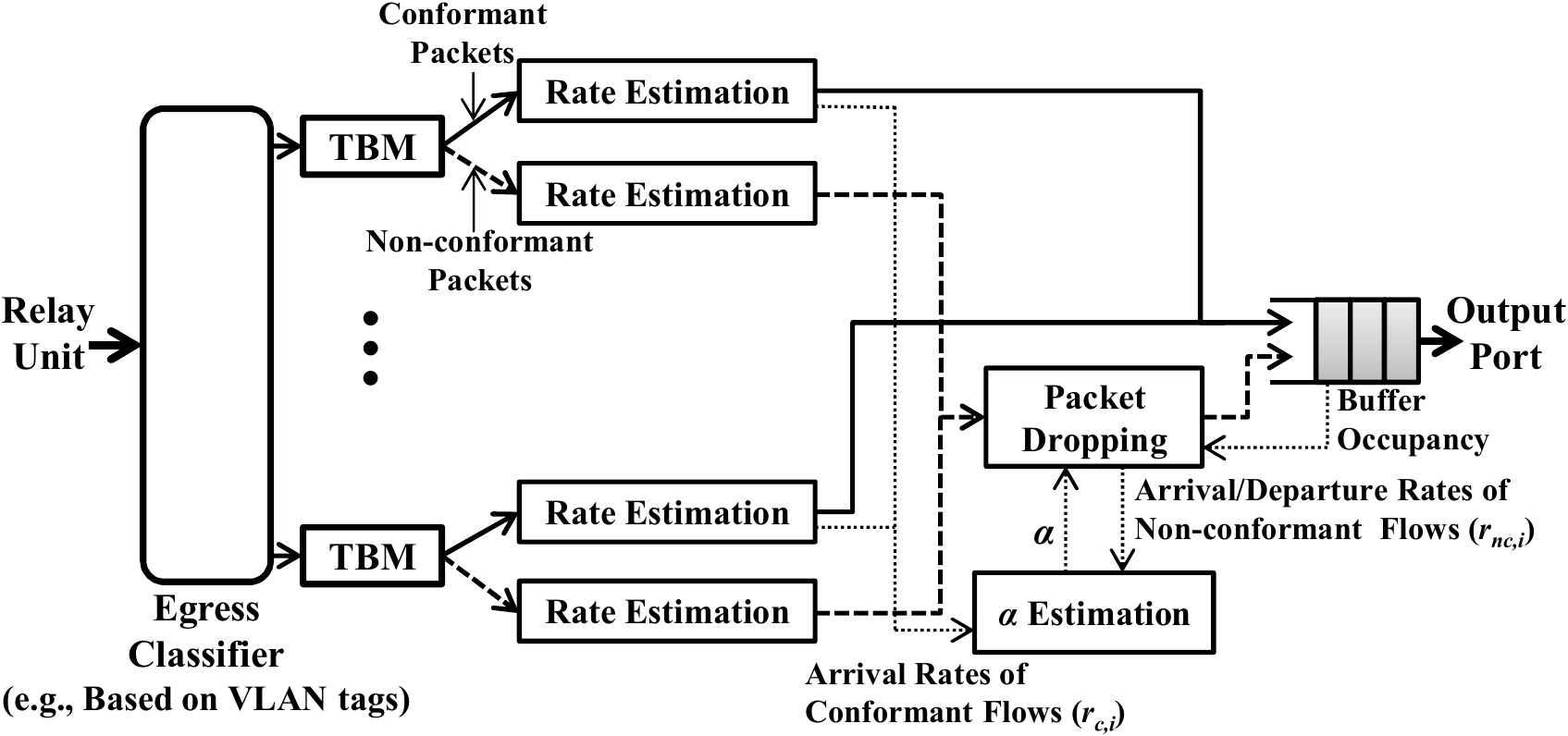} \\
{\scriptsize (b)}
\end{center}
\caption{ISP per-subscriber traffic control enabling proportional allocation of excess bandwidth:
    (a) A conceptual model based on WFQ and (b) a practical implementation based on CSFQ.}
\label{fg:isp_traffic_control}
\end{figure}

Fig. \ref{fg:access_switch} shows an access switch for a shared access network.
Considering the requirements in Sec. \ref{sec-2-1}, one can come up with a
conceptual model of ISP per-subscriber traffic control shown in
Fig. \ref{fg:isp_traffic_control} (a), which enables proportional allocation of
excess bandwidth based on weighted fair queueing (WFQ) and priority queueing
(PQ) with token bucket meters (TBMs): The first requirement is met by the use of
token bucket meters and PQ with higher priority for conformant packets, while
the second requirement is met by WFQ.  Note that, even with PQ ahead, WFQ can
still maintain its fairness property \cite{wang05:_packet}.

This conceptual model based on WFQ, however, has a major flaw: Due to the
separation of traffic from the same subscriber into two flows and separate
queueing, packet sequence is not preserved, which makes it impractical for user
datagram protocol (UDP) applications.  Fig. \ref{fg:isp_traffic_control} (b)
shows a practical implementation based on CSFQ, which can preserve packet
sequence through a common first in, first out (FIFO) queue. The architecture
shown in Fig. \ref{fg:isp_traffic_control} (b) corresponds to the extreme case
of CSFQ islands, i.e., the node itself is an island. Because both edge and core
router functionalities reside in the same node, there is no need to carrying
labels in packets between the rate estimation and the packet dropping units.

Let $A(t)$ be the total arrival rate of non-conformant packets at time $t$,
i.e., $A(t)\triangleq\sum_{i=1}^{N}r_{nc,i}(t)$,
where $r_{nc,i}(t)$ is the arrival rate of non-conformant packets for the
\(i\)th subscriber. If $A(t) > C_{ex}(t)$, the \emph{normalized} fair rate
$\alpha(t)$ is a unique solution to
\begin{equation}
C_{ex}(t) = \sum_{i=1}^{N} w_i \min(\alpha(t),~r_{nc,i}(t)/w_i) ,
\label{eq:fair_rate}
\end{equation}
where $w_i$ is the weight for the \(i\)th subscriber, which is proportional to
the token generation rate; otherwise, $\alpha(t)$ is set to
$max_i\left(r_{nc,i}(t)/w_i\right)$ \cite{stoica03:_core}. Based on the excess
bandwidth, arrival rates, and normalized fair rate, we can now implement rate
estimation and packet dropping as described in
Algorithm\(~\)\ref{alg:pseudocode_rate_drop}, which is a modified version of weighted
CSFQ with two arrival rates per subscriber:
If $r_{nc,i}(t)/w_{i}\leq\alpha(t)$, a non-conformant packet will be enqueued
for forwarding; otherwise, the packet will be dropped with the probability of
$\max\left(0,~1-\alpha(w_{i}/r_{i})\right)$.
%%% Customize algorithm environment
% %% C-style syntax
% \SetStartEndCondition{ (}{)}{)}\SetAlgoBlockMarkers{}{\}}%
% \SetKwProg{Fn}{}{\{}{}\SetKwFunction{FRecurs}{void FnRecursive}%
% \SetKwFor{For}{for}{\{}{}%
% \SetKwIF{If}{ElseIf}{Else}{if}{ \{}{elif}{else \{}{}%
% \SetKwFor{While}{while}{\{}{}%
% \SetKwRepeat{Repeat}{repeat\{}{until}%
% \AlgoDisplayBlockMarkers\SetAlgoNoLine%
%% compact style
\SetStartEndCondition{ }{}{}%
\SetKwProg{Fn}{}{\{}{}\SetKwFunction{FRecurs}{void FnRecursive}%
\SetKwFor{For}{for}{}{}%
% \SetKwIF{If}{ElseIf}{Else}{if}{}{elif}{else}{}%
\SetKwIF{If}{ElseIf}{Else}{if}{ then}{else if}{else}{endif}%
\SetKwFor{While}{while}{}{}%
\SetKwRepeat{Repeat}{repeat\{}{until}%
\AlgoDontDisplayBlockMarkers\SetAlgoNoEnd\SetAlgoNoLine%
\DontPrintSemicolon%
%% Functions
\SetKwFunction{CF}{Conform}%
\SetKwFunction{DR}{Drop}%
\SetKwFunction{EA}{Estimate$\alpha$}%
\SetKwFunction{EB}{ExcessBW}%
\SetKwFunction{EQ}{Enque}%
\SetKwFunction{ER}{EstimateRate}%
\SetKwFunction{LN}{Length}%
\SetKwFunction{MN}{Min}%
\SetKwFunction{MX}{Max}%
\SetKwFunction{UR}{UniformRandom}%
%% Misc.
%\SetAlgoLined%
%%% End of Customization
\begin{algorithm}[!t]
\emph{\textbf{On receiving} packet $p$ for the $i$th subscriber\;}
\eIf{$\CF{p} == True$}{
    $r_{c} \leftarrow$ \ER{$r_{c}$, \LN{$p$}}\;
    \EQ{$p$}\;
    $C_{ex} \leftarrow$ \EB{$r_{c}$}\tcc*[r]{using (\ref{eq:ex_bw})}
}{
    $r_{nc,i} \leftarrow$ \ER{$r_{nc,i}$, \LN{$p$}}\;
%   $p.rate \leftarrow r_{nc,i}$\;
    $prob \leftarrow$ \MX{$0,~1-\alpha(w_{i}/r_{nc,i})$}\;
    \eIf{$prob >$ \UR{0, 1}}{
        $\alpha \leftarrow$ \EA{$r_{nc,i}/w_{i}$,\LN{$p$},True}\;
        \DR{$p$}\;
    }
    {
        $\alpha \leftarrow$ \EA{$r_{nc,i}/w_{i}$,\LN{$p$},False}\;
        \EQ{$p$}\;
    }
}
\caption{Pseudocode of rate estimation and packet dropping.}
\label{alg:pseudocode_rate_drop}
\end{algorithm}

The estimation of the normalized fair rate (i.e., $\hat{\alpha}$ for $\alpha$)
is described in Algorithm\(~\)\ref{alg:pseudocode_fair_rate}, where $\hat{A}$
and $\hat{F}$ are the estimated aggregate arrival rate and the estimated
aggregate rate of the accepted traffic of non-conformant packets, respectively,
and $K_{\alpha}$ is a window size to filter out the inaccuracies in rate
estimation.
\begin{algorithm}[!t]
\SetKwProg{Fn}{Function}{}{end}\SetKwFunction{FEstAlpha}{Estimate$\alpha$}%
\Fn(){\FEstAlpha{r, l, dropped}}{
    \KwData{
        $r$ is the normalized arrival rate,
        $l$ a packet length, 
        and $dropped$ a flag indicating whether the packet is dropped or not.
        $\hat{\alpha}$ and $r_{max}$ are initialized to $C_{ex}$ and 0, respectively.
    }
    \KwResult{$\hat{\alpha}$ (i.e., fair share rate) is returned.}
    $\hat{A} \leftarrow$ \ER{$\hat{A},~l$}\;
    \If{$dropped == False$}{
        $\hat{F} \leftarrow$ \ER{$\hat{F},~l$}\;
    }
    \eIf{$\hat{A} \geq C_{ex}$}{
        \eIf{$congested == False$}{
            $congested \leftarrow True$\;
            $start\_time \leftarrow current\_time$\;
            \lIf{$\hat{\alpha} == 0$}{$\hat{\alpha} \leftarrow$ \MN{$r,~C_{ex}$}}
%           \If{$\hat{\alpha} == 0$}{
%               $\hat{\alpha} \leftarrow$ \MN{$r,~C_{ex}$}\;
%           }
        }
        {
            \If{$currentt\_time > start\_time + K_{\alpha}$}{
                $\hat{\alpha} \leftarrow \hat{\alpha} C_{ex}/\hat{F}$\;
                $start\_time \leftarrow current\_time$\;
                \lIf{$\hat{\alpha} == 0$}{$\hat{\alpha} \leftarrow$ \MN{$r,~C_{ex}$}}
%               \If{$\hat{\alpha} == 0$}{
%                   $\hat{\alpha} \leftarrow$ \MN{$r,~C_{ex}$}\;
%               }
            }
        }
    }(\tcc*[f]{$\hat{A} < C_{ex}$})
    {
        \eIf{$congested == True$}{
            $congested \leftarrow False$\;
            $start\_time \leftarrow current\_time$\;
            $r_{max} \leftarrow 0$\;
        }
        {
            \eIf{$currentt\_time < start\_time + K_{\alpha}$}{
                $r_{max} \leftarrow$ \MX{$r_{max},~r$}\;
            }
            {
                $start\_time \leftarrow current\_time$\;
                $\hat{\alpha} \leftarrow r_{max}$,~$r_{max} \leftarrow 0$\;
            }
        }
    }
    \Return $\hat{\alpha}$\;
}
\caption{Pseudocode of fair rate estimation.}
\label{alg:pseudocode_fair_rate}
\end{algorithm}
The update of the estimator $\hat{\alpha}$ is based on linear approximation of
the function $F(\cdot)$, i.e., $\hat{\alpha_{new}} = \hat{\alpha_{old}} \times
C_{ex}/\hat{F}$.

As discussed in \cite{stoica03:_core}, we use exponential averaging to estimate
various rates, i.e., $r_{c}$, $r_{nc,i}$, $A$ and $F$, whose general formula
is given by $x_{new}=\left(1-e^{-T/K}\right)\frac{l}{T}+e^{-T/K}x_{old}$,
where $T$ is elapsed time since the last update, which means the interarrival
time of corresponding packet, and $K$ is an averaging constant ($K_{\alpha}$ for
$A$ and $F$).

To better support bursty, elastic traffic like that of transmission control
protocol (TCP), we can also implement buffer-based amendment as in
\cite{stoica03:_core}. When receiving a packet, we check the buffer level
against a predefined threshold. Every time the buffer level passes the
threshold, we decrease $\hat{\alpha}$ by a small percentage (9\% for the
simulation in this letter). Note that the major purpose of this amendment in the
current scheme is to prevent non-conformant traffic from hogging the buffer
space of the common FIFO queue at the expense of conformant traffic, unlike that
of the original CSFQ.
\section{Simulation Results}
\label{sec-3}

\begin{figure}[!t]
\centering
\includegraphics[angle=-90,width=.85\linewidth]{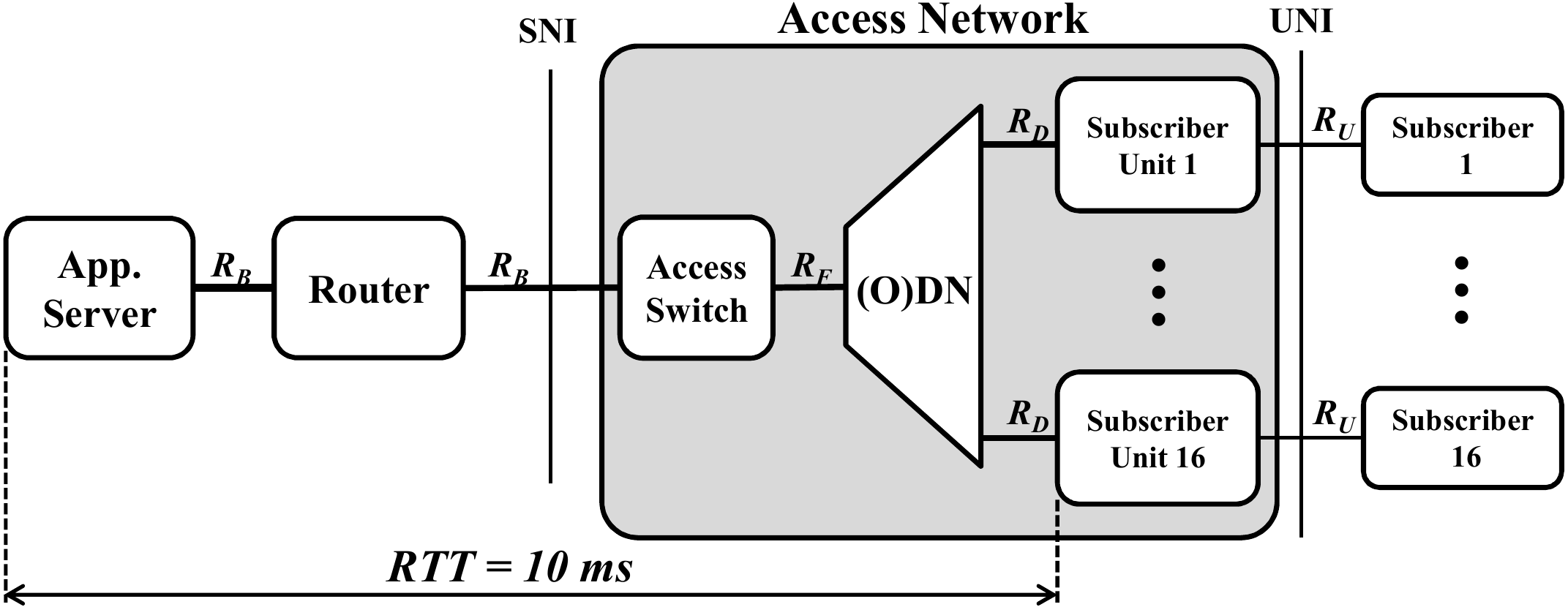}
\caption{A simulation model for a shared access network with 16 subscribers.}
\label{fg:simulation_model}
\end{figure}

We carried out a comparison study of the proposed scheme with the conceptual
model as a reference. For WFQ implementation in the conceptual model, we use
deficit round-robin (DRR)
\cite{Shreedhar:96-1}. Fig.\(~\)\ref{fg:simulation_model} shows a simulation
model where 16 subscribers are connected through 100-Mb/s user-network
interfaces (UNIs) to shared access with the same feeder and distribution rates
of 100-Mb/s, each of which receives packet streams from UDP or TCP sources in
the application server.
The backbone rate (i.e., $R_{B}$) and the end-to-end round-trip time are set to
10 Gb/s and 10 ms. To model the shared (optical) distribution network ((O)DN),
an Ethernet switch with the same feeder and distribution rates is used because
feeder and distribution links are identical (e.g., cable Internet) or passively
connected in a star topology (e.g., EPON) in shared access. The implementation
details are given in \cite{Kim:12-2}.

We divide 16 subscribers into 4 groups (i.e., 4 subscribers per group): For
Groups 1-3, each subscriber receives a 1000-byte packet at every 0.5 ms (i.e.,
the source rate of 16 Mb/s) from a UDP source. Token generation rate, however,
is set to 2.5 Mb/s for Group 1, 5 Mb/s for Group 2 and 7.5 Mb/s for Group 3. We
also set starting time to 0 s, 60 s, 120 s, respectively. For Group 4, each
subscriber receives packets from a greedy TCP source with token generation rate
of 10 Mb/s and starting time of 180 s. Token bucket size is set to 1 MB for all
subscribers, and peak rate control is not used at all. The size of FIFO and
per-subscriber queues of DRR is set to 1 MB (i.e., 17 MB in total) for the
reference scheme (denoted as ``DRR+TBM''), and the size of common FIFO queue is
set to 16 MB for the CSFQ-based scheme without (``CSFQ1+TBM'') and with
buffer-based amendment (``CSFQ2+TBM'') to cope with worst-case bursts resulting
from 16 token buckets with size of 1 MB each; as for the buffer-based amendment,
we set a threshold to 64 kB. The averaging constants used in the estimation of
flow rates (i.e, $K$) and the normalized fair rate (i.e., $K_{\alpha}$) are set
to 100 ms and 200 ms, respectively.

\begin{figure}[!tpb]
\begin{center}
\includegraphics[width=.8\linewidth,trim=35 15 5 42,clip=true]{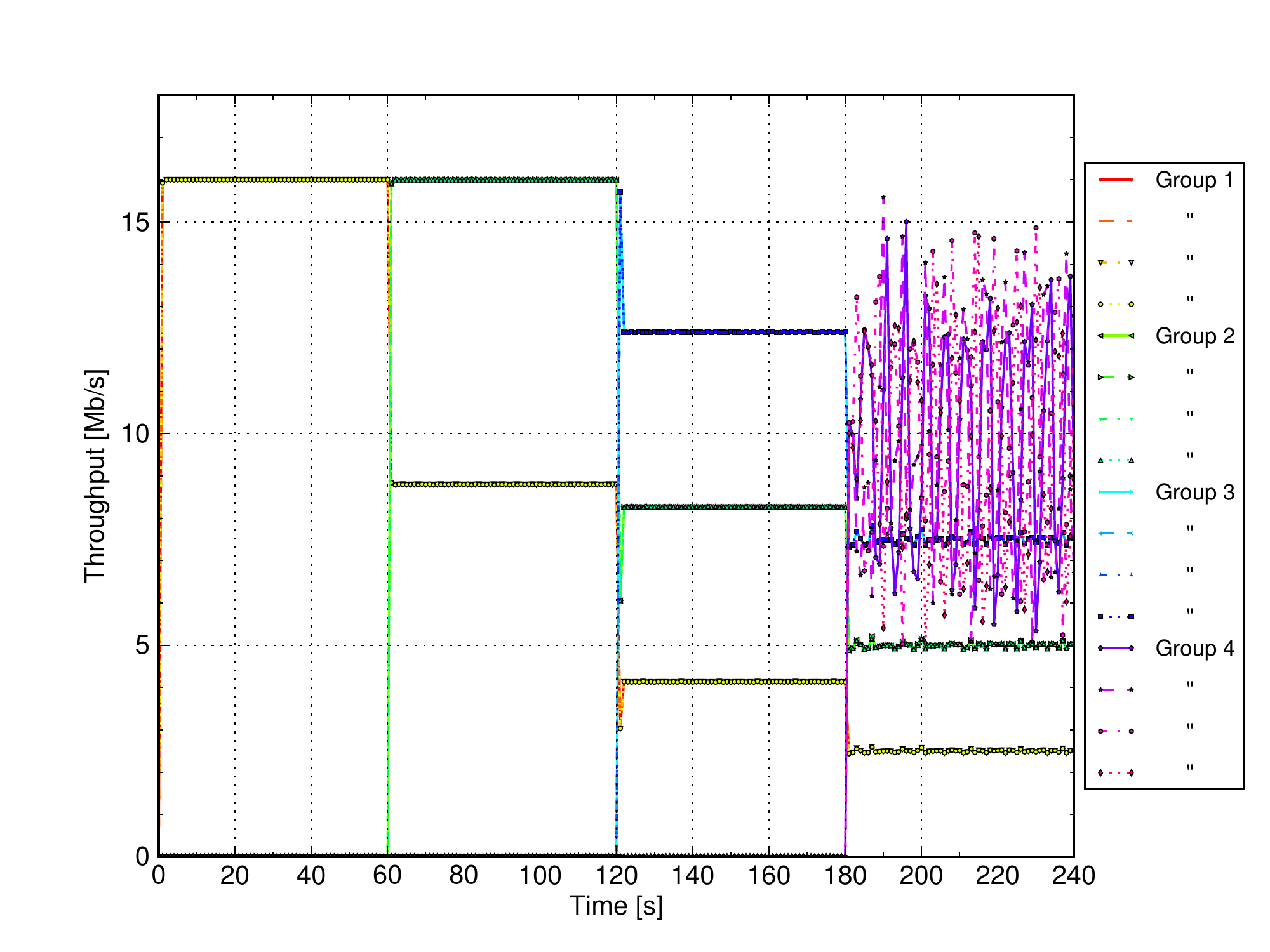}\\
{\scriptsize (a)}\\
\includegraphics[width=.8\linewidth,trim=35 15 5 25,clip=true]{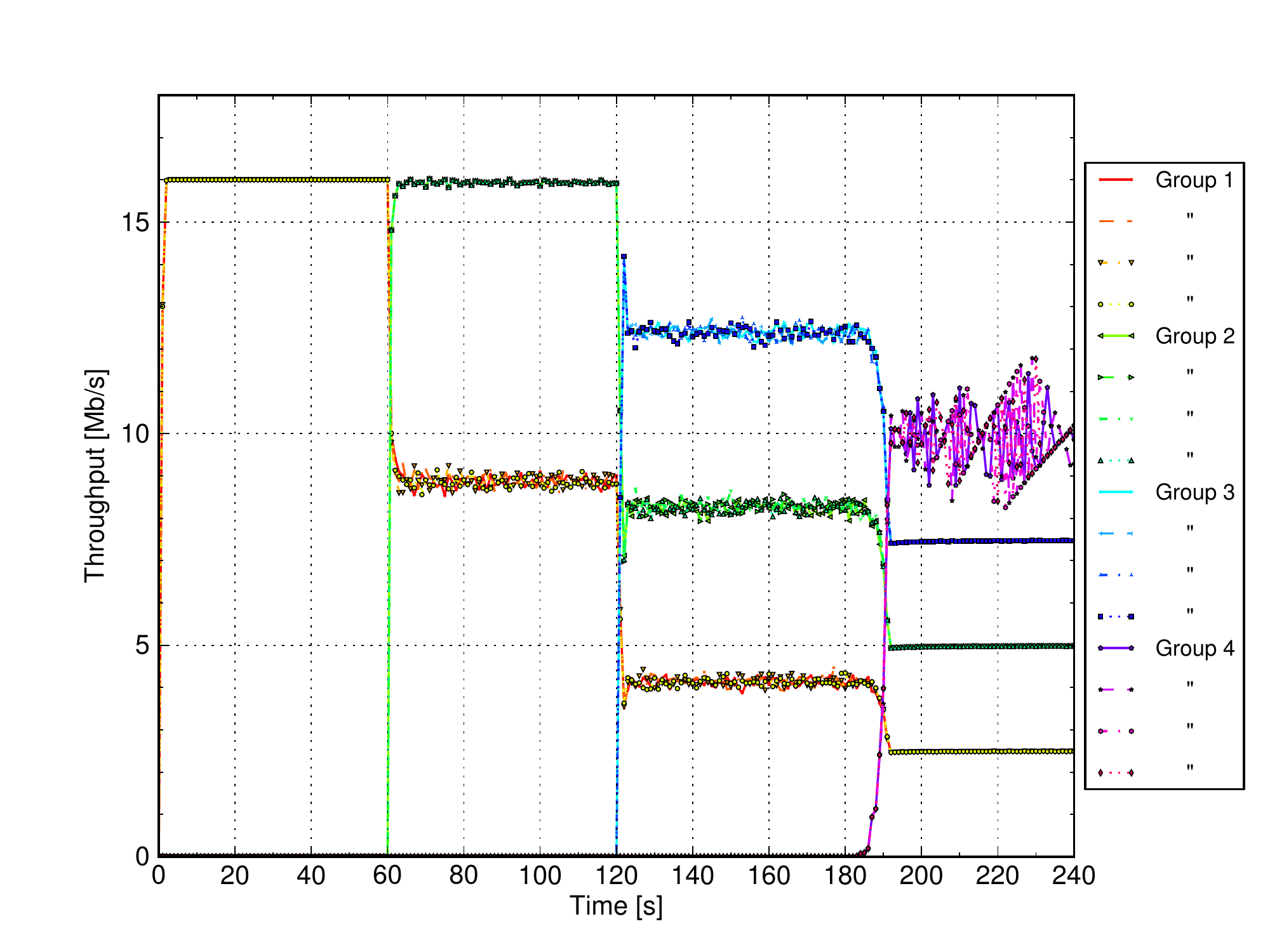}\\
{\scriptsize (b)}\\
\includegraphics[width=.8\linewidth,trim=35 15 5 25,clip=true]{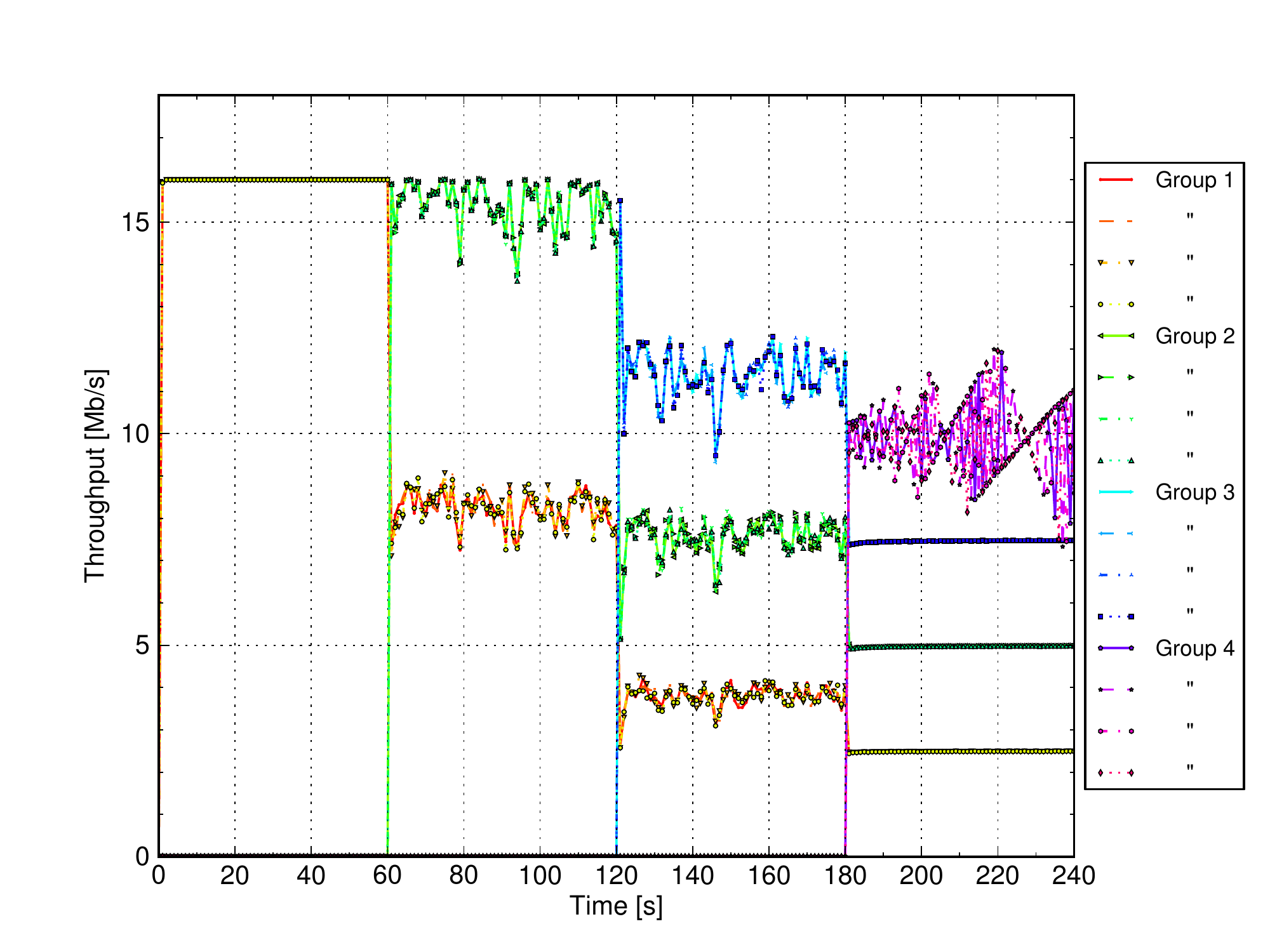}\\
{\scriptsize (c)}
\end{center}
\caption{Time series of throughput of flows: (a) DRR+TBM, (b) CSFQ1+TBM, and (c) CSFQ2+TBM.}
\label{fg:thruput_time}
\end{figure}

Fig. \ref{fg:thruput_time} shows flow throughput averaged over a 1-s interval
from one sample run, which demonstrates dynamic performances of each scheme
(i.e., \emph{how quickly} it can respond to the changes in incoming traffic and
allocate excess bandwidth accordingly). Until 180 s when TCP flows start, all
three schemes can allocate available bandwidth (including excess bandwidth)
among UDP flows well, with DRR+TBM being the best in terms of fluctuation and
convergence speed. Due to 1-MB token buckets, there are spikes in the throughput
of newly started flows at 60 s (i.e., Group 2) and 120 s (i.e., Group 3), while
the throughput of existing flows temporarily plunged accordingly. As TCP flows
start at 180 s, the difference among the three schemes become clearer: Because
packet sequence is not preserved in DRR+TBM, which causes lots of
retransmissions, throughput of TCP flows fluctuate most. With CSFQ1+TBM, while
the fluctuation in TCP flow throughput is not so big, the convergence is quite
slow (about 10 s to reach the token generation rate of 10 Mb/s). In this regard
we found that the buffer-based amendment in CSFQ2+TBM efficiently reduces the
transient period, especially for TCP flows, at the slight expense of
fluctuations in steady states.

\begin{figure}[!tpb]
\begin{center}
\includegraphics[width=.8\linewidth,trim=35 15 55 38,clip=true]{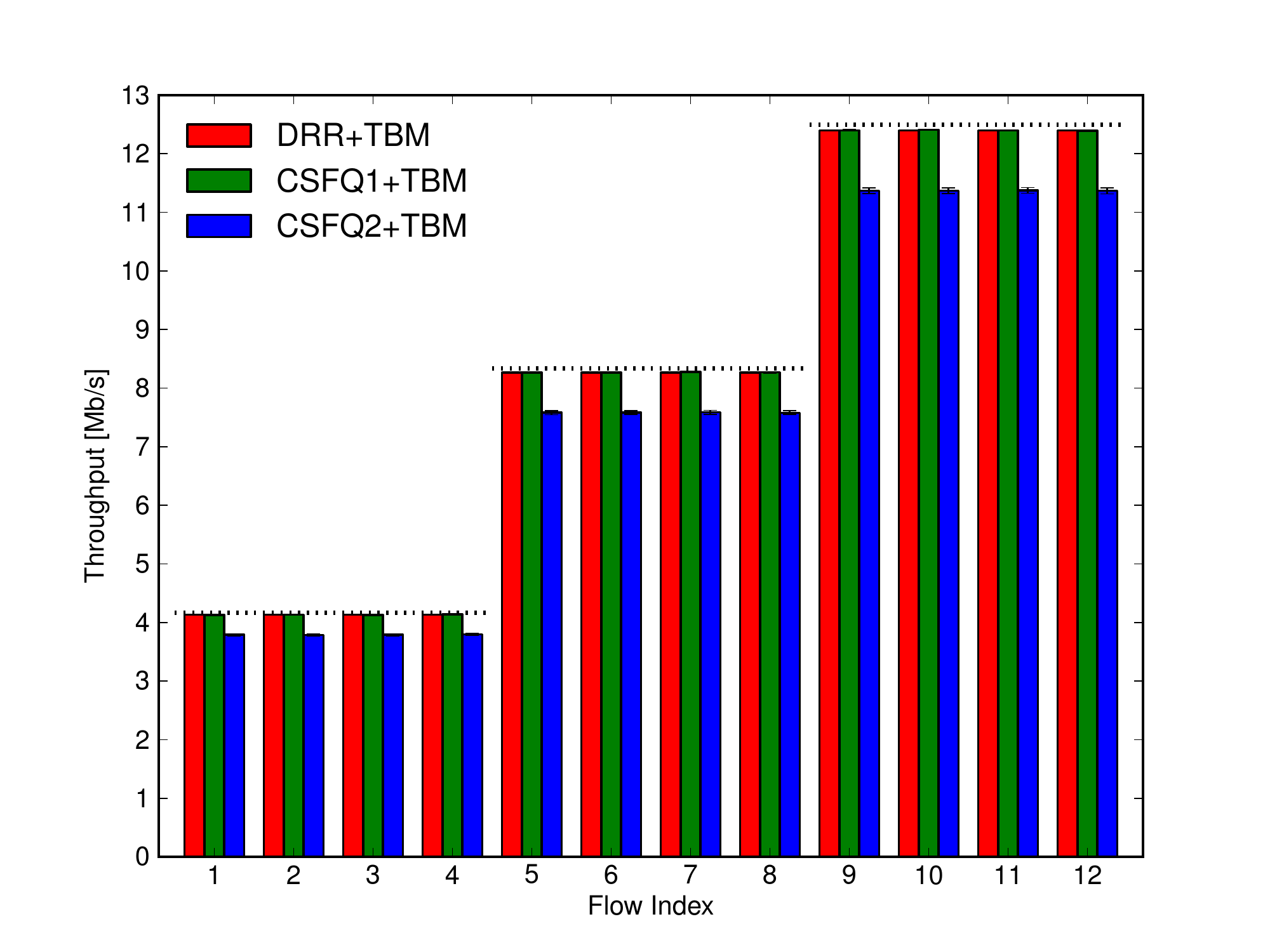}\\
{\scriptsize (a)}\\
\includegraphics[width=.8\linewidth,trim=35 15 55 25,clip=true]{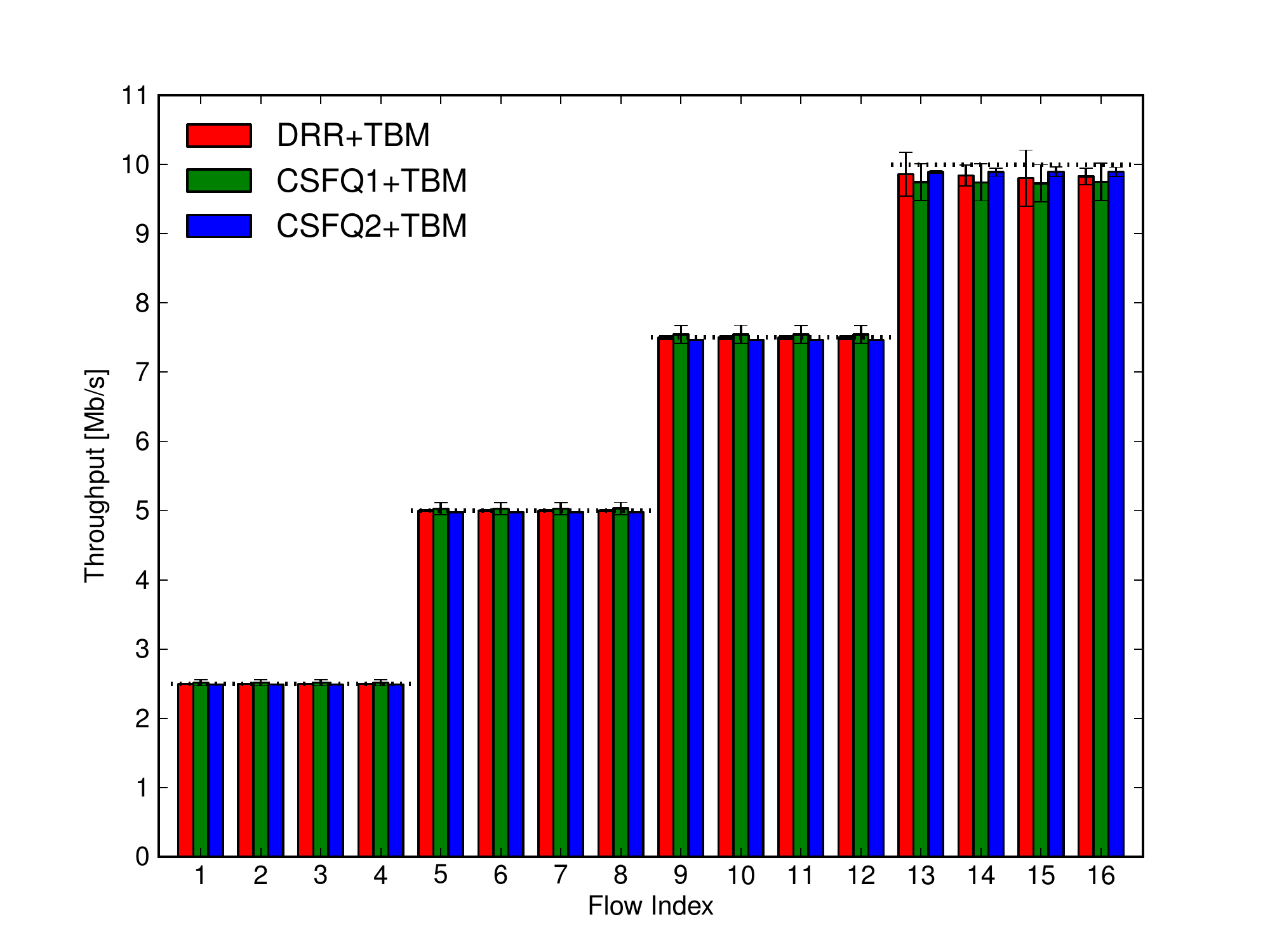}\\
{\scriptsize (b)}
\end{center}
\caption{Average throughput of flows with 95 percent confidence intervals for the period of (a) $130 \leq t < 180$ and (b) $190 \leq t < 240$.}
\label{fg:thruput_avg}
\end{figure}

Fig. \ref{fg:thruput_avg} shows the average throughput of flows for two 50-s
periods (i.e., a subperiod (60 s) minus a transient period (10 s)) with 95
percent confidence intervals from 10 repetitions, demonstrating static
performances of each scheme (i.e., \emph{how exactly} it can allocate available
bandwidth among subscribers per the requirements described in Sec. \ref{sec-2-1}
in a steady state). As shown in Fig. \ref{fg:thruput_avg} (a), both DRR+TBM and
CSFQ1+TBM allocate excess bandwidth from Group 4 exactly per
(\ref{eq:fair_rate}), while CSFQ2+TBM suffers from the fluctuations observed in
Fig. \ref{fg:thruput_time} (c). With TCP flows, however, CSFQ2+TBM performs best
and guarantees well the negotiated long-term average rates for newly started TCP
flows, even though the difference among the schemes is not that big. Note that
dotted lines indicate the fair share of each flow.
\section{Conclusions}
\label{sec-4}

In this letter we have studied the long-term aspect of resource sharing in ISP
traffic control for shared access networks and proposed ISP traffic control
schemes based on CSFQ and token bucket meters. Simulation results demonstrate
that the proposed schemes allocate excess bandwidth among active subscribers in
a fair and efficient way, while not compromising the service contracts specified
by the token bucket algorithm for conformant subscribers. With buffer-based
amendment, we could reduce transient period of the proposed scheme and thereby
improve throughput of interactive TCP flows.


\end{document}